\documentclass[oneside]{elsart}
\usepackage{graphicx}
\usepackage{amsmath}

\begin{document}
\begin{frontmatter}
\title{Phase diagrams of soluble multi-spin glass models}
\author{P. T. Muzy, A. P. Vieira, and S. R. Salinas}
\address{Instituto de F\'{\i}sica, 
Universidade de S\~{a}o Paulo\\
Caixa Postal 66318
05315-970, S\~{a}o Paulo, SP, Brazil}
\date{12 April 2005}

\begin{abstract}
We include $p$-spin interactions in a spherical version of a soluble
mean-field spin-glass model proposed by van Hemmen. Due to the simplicity of
the solutions, which do not require the use of the replica trick, we are
able to carry out a detailed investigation of a number of special
situations. For $p\geq 3$, there appear first-order transitions between the
paramagnetic and the ordered phases. In the presence of additional
ferromagnetic interactions, we show that there is no stable mixed phase,
with both ferromagnetic and spin-glass properties.
\end{abstract}
\end{frontmatter}

\section{Introduction}

The attempts to account for the physical properties of disordered systems
have been a continuous source of models and problems in statistical
mechanics. It is known that a mean-field $p$-spin-glass model, with $p\geq 3$%
, displays a first-order transition to a glassy phase, and non-trivial
dynamical behavior. Although models with $p$-spin interactions may be far
from representing real glasses, it has been noted that the dynamical
mean-field equations do resemble the corresponding mode-coupling expressions
for glassy systems \cite{cugliandolo}.

In some recent publications, a ferromagnetic term, as well as additional
multi-spin interactions, have been added to the simple spin-glass $p$-spin
model \cite{hertz99,gillin00,crisanti04}. These new systems
display continuous and first-order transitions, and a glassy ferromagnetic
region with pronounced reentrant borders. However, the search for the
thermodynamic solutions of these mean-field models requires the use of the
subtleties of the replica method. The replica-symmetric solutions have to be
supplemented by a one-step replica-symmetry breaking scheme in order to
characterize the glassy ferromagnetic phase.

In this paper, we decided to investigate the thermodynamic behavior of
highly simplified versions of disordered multi-spin mean-field models. The
idea consists in the inclusion of $p$-spin interactions, with $p\geq 3$, in
a slightly more general version of a spin-glass model proposed by van Hemmen
about twenty years ago \cite{vanhemmen82}. This simple van Hemmen model,
which does not require the use of the replica method, is obviously unable to
reproduce the rich (ultrametric) structure of the spin-glass phase \cite%
{choy}. However, it does give a reasonable account of the main features of
the phase diagrams, including the location of phase regions and transition
lines. In this paper, we further simplify the calculations by assuming
spherical instead of Ising spin variables.

In Section 2, we review the van Hemmen spin-glass model with spherical spin
variables. In Section 3, we introduce multi-spin interactions in the van
Hemmen model, and make a number of contacts with recent work. Section 4 is
devoted to some conclusions.

\section{The van Hemmen model}

The van Hemmen spin-glass model, with the inclusion of a ferromagnetic term,
is given by the Hamiltonian 
\begin{equation}
H=-\frac{J_{0}}{N}\sum_{1\leq i<j\leq N}\sigma _{i}\sigma _{j}-\sum_{1\leq
i<j\leq N}J_{ij}\sigma _{i}\sigma _{j}-H\sum_{i=1}^{N}\sigma _{i},
\label{101}
\end{equation}
where $J_{0}>0$, 
\begin{equation}
J_{ij}=\frac{1}{2N}J\left[ \xi _{i}^{\left( 1\right) }\xi _{j}^{\left(
2\right) }+\xi _{i}^{\left( 2\right) }\xi _{j}^{\left( 1\right) }\right] ,
\label{102}
\end{equation}
$J>0$, and $H$ is an external field. The set of independent, identically
distributed random variables $\left\{ \xi _{i}^{\left( \alpha \right)
}\right\} $, for $i=1,2,...,N$ and $\alpha =1,2$, associated with a
double-delta probability distribution, 
\begin{equation}
p\left( \xi _{i}^{\left( \alpha \right) }\right) =\frac{1}{2}\delta \left(
\xi _{i}^{\left( \alpha \right) }-1\right) +\frac{1}{2}\delta \left( \xi
_{i}^{\left( \alpha \right) }+1\right) ,  \label{103}
\end{equation}
is supposed to mimic the presence of disorder and competition in real spin
glasses. In the spherical version of this model, the spin variables are real
numbers, $-\infty \leq \sigma _{i}\leq \infty $, for all $i=1,2,...,N$, with
the spherical constraint, 
\begin{equation}
\sum_{i=1}^{N}\sigma _{i}^{2}=N.
\end{equation}

Using more convenient variables, 
\begin{equation}
m=\frac{1}{N}\sum_{i=1}^{N}\sigma _{i},\qquad m_{\alpha }=\frac{1}{N}%
\sum_{i=1}^{N}\xi _{i}^{\left( \alpha \right) }\sigma _{i},  \label{105}
\end{equation}
for $\alpha =1,2$, and discarding irrelevant terms in the thermodynamic
limit, we can rewrite the Hamiltonian, given by Eq. (\ref{101}), in the form 
\begin{equation}
H=-\frac{1}{2}J_{0}Nm^{2}-\frac{1}{2}JNm_{1}m_{2}-HNm.  \label{107}
\end{equation}
Given a configuration of the random variables, the partition function may be
written as 
\begin{equation}
Z=\mathrm{Tr}\exp \left[ \frac{1}{2}\beta J_{0}Nm^{2}+\frac{1}{2}\beta
JNm_{1}m_{2}+\beta HNm\right] ,  \label{108}
\end{equation}
where $\beta $ is the inverse of temperature, and the trace should take into
account the spherical constraint.

We now rewrite the partition function in the form 
\begin{eqnarray}
Z & = &\int_{-\infty }^{+\infty }dm\int_{-\infty }^{+\infty }dm_{1}\int_{-\infty
}^{+\infty }dm_{2}\Omega \left( m,m_{1},m_{2}\right) \nonumber \\
& \times & \exp \left\{\frac{1}{2}\beta 
J_{0}Nm^{2} +
\frac{1}{2}\beta JNm_{1}m_{2}+\beta HNm\right\},
\end{eqnarray}
where 
\begin{eqnarray}
\Omega &= \mathrm{Tr}\prod_{i=1}^{N}\left( \int_{-\infty }^{+\infty }d\sigma _{i}\right) 
\delta \left( \sum_{i=1}^{N}\sigma _{i}^{2}-N\right) \delta \left( m-\frac{1%
}{N}\sum_{i=1}^{N}\sigma _{i}\right) \nonumber \\
& \times \delta \left( m_{1}-\frac{1}{N}\sum_{i=1}^{N}\xi _{i}^{\left(
1\right) }\sigma _{i}\right) \delta \left( m_{2}-\frac{1}{N}%
\sum_{i=1}^{N}\xi _{i}^{\left( 2\right) }\sigma _{i}\right) .
\end{eqnarray}
In the thermodynamic limit, we use an integral representation for the delta
functions and invoke the law of large numbers in order to obtain the
asymptotic result \cite{haddad04} 
\begin{equation}
\Omega \sim \exp \left\{ \frac{N}{2}\left[ 1+\ln \left( 2\pi \right) +\ln
\left( 1-m^{2}-m_{1}^{2}-m_{2}^{2}\right) \right] \right\} ,
\end{equation}
which leads to the partition function 
\begin{equation}
Z\sim \int_{-\infty }^{+\infty }dm\int_{-\infty }^{+\infty
}dm_{1}\int_{-\infty }^{+\infty }dm_{2}\exp \left[ -\beta Nf\left(
m,m_{1},m_{2}\right) \right] ,  \label{109}
\end{equation}
where 
\begin{equation}
f =-\frac{1}{2}\left(J_{0}m^{2}+Jm_{1}m_{2}\right)-Hm \\
 - \frac{1}{2\beta }\left[ 1+\ln \left( 2\pi \right) +\ln \left(
1-m^{2}-m_{1}^{2}-m_{2}^{2}\right) \right] .  \label{110}
\end{equation}
The thermodynamic solutions come from the stable minima of this free-energy
functional.

In zero external field, $H=0$, it is easy to draw the phase diagram of
Figure 1, in terms of $T=\left( \beta J\right) ^{-1}$ and $r=J_{0}/J$.
Besides the disordered paramagnetic phase ($m=m_{1}=m_{2}=0$), there is a
``spin-glass'' ($m=0$, and $m_{1},m_{2}\neq 0$) and a ferromagnetic phase ($%
m\neq 0$, and $m_{1}=m_{2}=0$). In the ferromagnetic region, $m^{2}=1-T/r$;
in the spin-glass region, $m_{1}^{2}=m_{2}^{2}=1/2-T$. The paramagnetic
borders are lines of continuous phase transitions. Mixed phases (with $m\neq
0$ and $m_{1},m_{2}\neq 0$) are restricted to a line of coexistence at $%
r=J_{0}/J=1/2$ (which also corresponds to the limits of stability of the
ferromagnetic and spin-glass solutions). Similar topological features are
also present in the phase diagram of the much more elaborate
Sherrington-Kirkpatrick (SK) mean-field model of a spin glass.

\begin{figure}
\begin{center}
\includegraphics[width=.6\columnwidth]{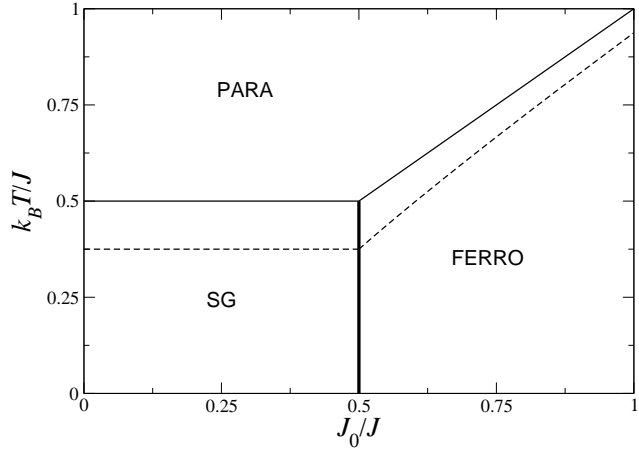}
\end{center}
\caption{Phase diagram of the spherical van Hemmen model. Solid lines correspond to zero external field, while dashed lines are obtained for $H_{R}/J=1/4$. Thin (thick) lines indicate second (first) order transitions. The labels correspond to the paramagnetic (PARA), ferromagnetic (FERRO), and to the ``spin-glass'' (SG) phases.}
\end{figure}

In the presence of a random field, the Hamiltonian of the van Hemmen model
is written as 
\begin{equation}
H=-\frac{J_{0}}{N}\sum_{1\leq i<j\leq N}\sigma _{i}\sigma _{j}-\sum_{1\leq
i<j\leq N}J_{ij}\sigma _{i}\sigma _{j}-\sum_{i=1}^{N}H_{i}\sigma _{i},
\label{1101}
\end{equation}
where $\left\{ H_{i}\right\} $, for $i=1,2,...,N$, is a set of quenched
independent, identically distributed random variables, given by the
probability distribution 
\begin{equation}
p_{H}\left( H_{i}\right) =\frac{1}{2}\delta \left( H_{i}-H_{R}\right) +\frac{%
1}{2}\delta \left( H_{i}+H_{R}\right) .  \label{1102}
\end{equation}
Introducing the new variable 
\begin{equation}
q=\frac{1}{N}\sum_{i=1}^{N}\frac{H_{i}}{H_{R}}\sigma _{i},  \label{1103}
\end{equation}
it is straightforward to obtain the free-energy functional 
\begin{equation}
f =-\frac{1}{2}\left(J_{0}m^{2}+Jm_{1}m_{2}\right)-H_{R}q \\
 - \frac{1}{2\beta }\left[ 1+\ln \left( 2\pi \right) +\ln \left(
1-m^{2}-m_{1}^{2}-m_{2}^{2}\right) \right] .  \label{1104}
\end{equation}
In the $T-r$ phase diagram, there is a depression of the paramagnetic lines,
which still meet at $r=1/2$ (see Figure 1). For small values of $%
h_{R}=H_{R}/J<<1$, we have the following asymptotic forms of the
paramagnetic critical lines: (i) $T=1/2-2h_{R}^{2}$, at the spin-glass
border; (ii) $T=r-h_{R}^{2}/r$, at the paramagnetic-ferromagnetic border.
There is no tricritical point with spherical spin variables. The same
qualitative depression of the paramagnetic borders is also present in the
phase diagram of an SK model in a random field \cite{soares}.

\section{Multi-spin interactions}

The inclusion of $p$-spin interactions, with $p\geq 2$, leads to a natural
generalization of the van Hemmen model. Let us write the spin Hamiltonian 
\begin{equation}
H=-\sum_{1\leq i_{1}<i_{2}<...<i_{p}\leq N}J_{i_{1}...i_{p}}\sigma
_{i_{1}}...\sigma _{i_{p}},
\end{equation}
with 
\begin{equation}
J_{i_{1}...i_{p}}=\frac{J_{p}}{p!N^{p-1}}\sum_{\left( \alpha _{1}...\alpha
_{p}\right) }\xi _{i_{1_{{}}}}^{\left( \alpha _{1}\right) }\xi
_{i_{2_{{}}}}^{\left( \alpha _{2}\right) }...\xi _{i_{p_{{}}}}^{\left(
\alpha _{p}\right) },
\end{equation}
where the sum is over the permutations of the $p$ patterns, $\alpha
_{1},...,\alpha _{p}$. In analogy with the $p=2$ case, we have enlarged the
set of random variables, $\left\{ \xi _{i}^{\left( \alpha \right) }\right\} $%
, which are still given by the probability distribution of Eq. (\ref{103}).
With this choice of patterns, the $p$-spin Hamiltonian can be written as 
\begin{equation}
H=-\frac{1}{p!}J_{p}Nm_{1}...m_{p},
\end{equation}
from which we obtain the free-energy functional 
\begin{equation}
f=-\frac{1}{p!}J_{p}m_{1}...m_{p}-\frac{1}{2\beta }\left[ 1+\ln \left( 2\pi
\right) +\ln \left( 1-\sum_{\alpha =1}^{p}m_{\alpha }^{2}\right) \right] .
\end{equation}
For $p\geq 3$, it is easy to show that there is a first-order transition
between a paramagnetic disordered phase ($m_{1}=m_{2}=...=0$) and an ordered
``spin-glass''\ phase ($m_{1}^{2}=m_{2}^{2}=...\neq 0$). The existence of a
first-order transition in these $p$-spin models, which leads to spinodal
lines and may be responsible for peculiar dynamical phenomena, is usually
taken as an important contact with the behavior of real glasses.

These calculations can be easily extended to a much larger class of models.
Let us consider blocks of $p$-spin and $r$-spin interactions, as well as
ferromagnetic $n$-spin interactions and a random external field. A
sufficiently general spin Hamiltonian may be written as 
\begin{equation*}
H=-\sum_{1\leq i_{1}<i_{2}<...<i_{p}\leq N}J_{i_{1}...i_{p}}\sigma
_{i_{1}}...\sigma _{i_{p}}-\sum_{1\leq i_{1}<i_{2}<...<i_{r}\leq
N}J_{i_{1}...i_{r}}\sigma _{i_{1}}...\sigma _{i_{r}}-
\end{equation*}%
\begin{equation}
-\frac{J_{0}}{n!N^{n-1}}\left[ \sum_{i=1}^{N}\sigma _{i}\right]
^{n}-\sum_{i=1}^{N}H_{i}\sigma _{i},  \label{201}
\end{equation}%
with 
\begin{equation}
J_{i_{1}...i_{p}}=\frac{J_{p}}{p!N^{p-1}}\sum_{\left( \alpha _{1}...\alpha
_{p}\right) }\xi _{i_{1_{{}}}}^{\left( \alpha _{1}\right) }\xi
_{i_{2_{{}}}}^{\left( \alpha _{2}\right) }...\xi _{i_{p_{{}}}}^{\left(
\alpha _{p}\right) },  \label{202}
\end{equation}%
and 
\begin{equation}
J_{i_{1}...i_{r}}=\frac{J_{r}}{r!N^{r-1}}\sum_{\left( \alpha _{p+1}...\alpha
_{p+r}\right) }\xi _{i_{1_{{}}}}^{\left( \alpha _{p+1}\right) }\xi
_{i_{2_{{}}}}^{\left( \alpha _{p+2}\right) }...\xi _{i_{r_{{}}}}^{\left(
\alpha _{p+r}\right) },  \label{203}
\end{equation}%
where the sums are over all permutations of $\alpha _{1}...\alpha _{p}$ and $%
\alpha _{p+1}...\alpha _{p+r}$, and the independent random variables are
still given by the probability distributions of equations (\ref{103}) and (%
\ref{1102}).

Using the definitions of $m$, $m_{\alpha }$, and $q$, given by Eqs. (\ref%
{105}) and (\ref{1103}), for $\alpha =1,...,p+r$, it is easy to rearrange
the Hamiltonian in the more convenient form 
\begin{equation}
H=-\frac{1}{p!}J_{p}Nm_{1}...m_{p}-\frac{1}{r!}NJ_{r}m_{p+1}...m_{p+r}-\frac{%
1}{n!}J_{0}Nm^{n}-H_{R}Nq.
\end{equation}%
In the thermodynamic limit, assuming finite values of the parameters $p$, $r$%
, and $n$, we can write the free-energy functional 
\begin{eqnarray}
f&=&-\frac{1}{p!}J_{p}m_{1}...m_{p}-\frac{1}{r!}J_{r}m_{p+1}...m_{p+r}-\frac{1%
}{n!}J_{0}m^{n}-H_{R}q \nonumber \\
&-&\frac{1}{2\beta }\left[ 1+\ln \left( 2\pi \right) +\ln \left(
1-m^{2}-q^{2}-\sum_{\alpha =1}^{p+r}m_{\alpha }^{2}\right) \right] .
\label{205}
\end{eqnarray}

We now consider some particular situations:\bigskip

\begin{figure}
\begin{center}
\includegraphics[width=.6\columnwidth]{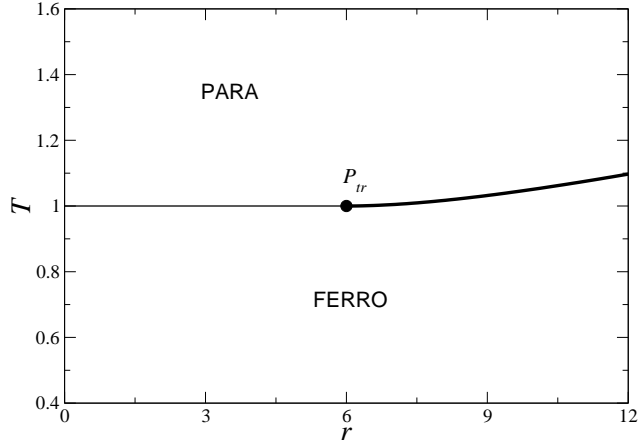}
\end{center}
\caption{Phase diagram of the mean-field spherical model with ferromagnetic $J_{02}$ and $J_{04}$ couplings in zero external field. Thin (thick) lines indicate second (first) order transitions, while $P_{tr}$ marks the tricritical point.}
\end{figure}

\begin{figure}
\begin{center}
\includegraphics[width=.7\columnwidth]{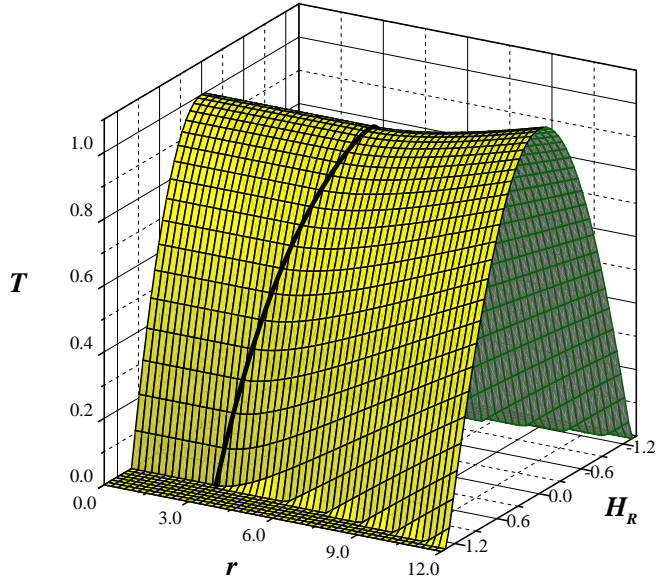}
\end{center}
\caption{Phase diagram of the mean-field spherical model with ferromagnetic $J_{02}$ and $J_{04}$ couplings in a random field of strength $H_R$. The surface corresponds to a paramagnetic boundary, and the thick curve is a line of tricritical points, separating a region of first-order (large $r$) from a region of second-order (small $r$) phase transitions.}
\end{figure}

(A) A model with (uniform) ferromagnetic interactions $J_{02}$ and $J_{04}$ involving blocks of $%
n=2$ and $n=4$ spins, in the presence of a random field, leads to the
free-energy functional 
\begin{equation}
f_{u}=-\frac{1}{2}J_{02}m^{2}-\frac{1}{4!}J_{04}m^{4}-qH_{R}-\frac{1}{2\beta }%
\left[ 1+\ln \left( 2\pi \right) +\ln \left( 1-m^{2}-q^{2}\right) \right] .
\label{206}
\end{equation}%
In zero random field, $H_{R}=0$, the phase diagram, in terms of temperature, 
$T=\left( \beta J_{02}\right) ^{-1}$, versus the ratio of ferromagnetic
uniform interactions, $r=J_{04}/J_{02}$, displays a line of second-order
transitions that turns into a first-order boundary at a tricritical point, $%
J_{04}/J_{02}=6$ (see Figure 2). The existence of first and second-order
transitions at the mean-field level gives an additional motivation to pursue
the investigation of the spin-glass versions of this model. The random field
introduces just a shift of the tricritical point. The global phase diagram,
in terms of $T$, $J_{04}/J_{02}$, and the strength of the random field, $%
H_{R}/J_{02}$, as shown in Figure 3, displays a line of tricritical points.

(B) If we assume $r=0$ and $n=2$, in zero field, it is possible to make
contact with some recent calculations for a spherical version of a $p$-spin
Sherrington-Kirkpatrick (SK) spin-glass model with the addition of a simple
ferromagnetic term. According to these calculations \cite{hertz99,gillin00}, for $p\geq 3$, the phase diagram of the spherical $p$-spin SK
model displays a glassy ferromagnetic phase, besides the usual paramagnetic,
spin glass, and pure ferromagnetic phases. Also, there is a pronounced
reentrance of the border between the two ferromagnetic phases.

In the context of the van Hemmen models, we write the free-energy functional 
\begin{equation}
f = -\frac{1}{p!}J_{p}m_{1}...m_{p}-\frac{1}{2}J_{0}m^{2}
- \frac{1}{2\beta }\left[ 1+\ln \left( 2\pi \right) +\ln \left(
1-m^{2}-\sum_{\alpha =1}^{p}m_{\alpha }^{2}\right) \right] .  \label{207}
\end{equation}%
We now set $p=3$, which gives typical results for $p>n$. The minimization of 
$f$ leads to the phase diagram shown in Figure 4, in terms of temperature, $%
T=1/\left( \beta J_{3}\right) $, and the ratio of interactions, $%
r=J_{0}/J_{3}$. The paramagnetic phase ($m=m_{1}=m_{2}=m_{3}=0$) is stable
for $T>r$. The ferromagnetic phase ($m^{2}=1-T/r$, $m_{1}=m_{2}=m_{3}=0$) is
stable for $T<r$. The critical line between the paramagnetic and the
ferromagnetic phases ends at a critical endpoint (at $T=r=0.03278...$). The
\textquotedblleft spin-glass phase\textquotedblright\ ($m=0$, $%
m_{1}^{2}=m_{2}^{2}=m_{3}^{2}\neq 0$) is limited by a first-order boundary
(at $T=0.03278...$, for small values of $r$; and at $r=1/9\sqrt{3}$ for $T=0$%
). The extremization of the free energy $f$ also leads to a
\textquotedblleft mixed ferromagnetic solution\textquotedblright , $m\neq 0$%
, with $m_{1}^{2}=m_{2}^{2}=m_{3}^{2}\neq 0$, which might be the analogue of
the glassy ferromagnetic phase of the $p$-spin SK models. However, it is
easy to show that the free energy of this solution is always larger than the
free energy of the uniform ferromagnetic phase. For this $p=3$ example, at
any temperature, $f\left( \text{ferro}\right) -f\left( \text{mixed ferro}%
\right) =-18r^{3}<0$. Indeed, it is easy to obtain analogous results for $p>3$.

\begin{figure}
\begin{center}
\includegraphics[width=.6\columnwidth]{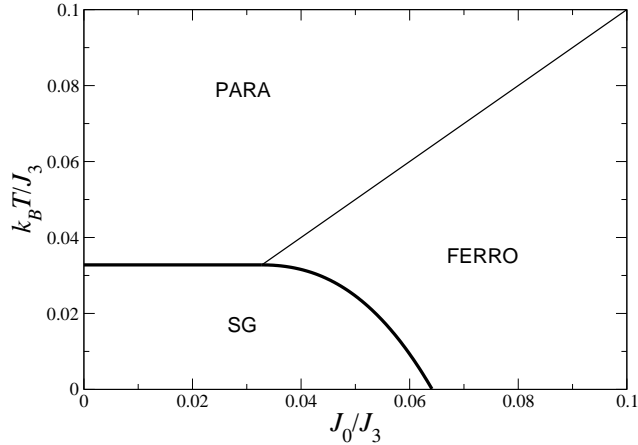}
\end{center}
\caption{Phase diagram of the spherical van Hemmen model with ferromagnetic
pair interactions and random $p=3$ couplings. Thin (thick) lines indicate second (first) order transitions.}
\end{figure}

\section{Conclusions}

We have introduced multi-spin interactions in the van Hemmen spin-glass
model with spherical spin variables. Using standard techniques, we write
analytic expressions for a self-averaged free energy of a model Hamiltonian
including blocks of disordered and uniform interactions, in the presence of
random fields. From this free energy, it is easy to draw some phase
diagrams, in order to compare with results for the corresponding SK models.

In particular, for $p=3$ and two-spin uniform ferromagnetic interactions, we
obtain a phase diagram with first-order transitions between spin-glass and
either paramagnetic or ferromagnetic phases, and a second-order
ferro-paramagnetic transition line, which ends at a critical endpoint. In
contrast to the calculations for the analogous SK model, we show that there
is no possibility of appearance of a ferromagnetic phase of spin-glass
character.


\begin{thebibliography}{9}
\bibitem{cugliandolo} L. Cugliandolo, Dynamics of glassy systems, in \emph{Slow Relaxations and Nonequilibrium Dynamics in Condensed Matter}, vol. 77 of \emph{Les Houches -- \'Ecole d'Et\'e de Physique Th\'eorique}, edited by J. Barrat, M. V. Feigelman, J. Kurchan, and J. Dalibard (Springer, New York, 2003); also online at arXiv: cond-mat/0210312.

\bibitem{hertz99} J. A. Hertz, David Sherrington, and Th. M. Nieuwenhuizen,
Phys. Rev. E\textbf{60}, R2460 (1999).

\bibitem{gillin00} Peter Gillin and David Sherrington, J. Phys. A: Math.
Gen. \textbf{33}, 3081 (2000).

\bibitem{crisanti04} A. Crisanti and L. Leuzzi, Phys. Rev. Lett. \textbf{93}%
, 217203 (2004).

\bibitem{vanhemmen82} J. L. van Hemmen, Phys. Rev. Lett. \textbf{49}, 409
(1982); J. L. van Hemmen, A. C. D. van Enter, and J. Canisius, Z. Phys. B%
\textbf{50}, 311 (1983).

\bibitem{choy} T. C. Choy and D. Sherrington, J. Phys. C\textbf{17}, 739
(1984).

\bibitem{haddad04} T. A. S. Haddad, A. P. Vieira, and S. R. Salinas, Physica
A\textbf{342}, 76 (2004).

\bibitem{soares} R. F. Soares, F. D. Nobre, and J. R. L. De Almeida, Phys.
Rev. B\textbf{50}, 6151 (1994).
\end{thebibliography}
\end{document}